\documentclass[aps,twocolumn,superscriptaddress,floatfix,nolongbibliography,noeprint]{revtex4-2}
\usepackage[colorlinks=true, citecolor=cyan, urlcolor=blue]{hyperref}
\usepackage{algorithm}
\usepackage{algpseudocode}
\usepackage{blindtext}
\usepackage{bm}
\usepackage{dsfont}
\usepackage{braket}
\usepackage{hyperref}
\usepackage{amssymb}
\usepackage{amsmath}
\usepackage{graphicx}
\usepackage{float}
\usepackage{bbold}
\usepackage{xcolor}
\usepackage{mathtools}

\renewcommand{\t}[1]{\textrm{#1}}
\newcommand{\RNum}[1]{\uppercase\expandafter{\romannumeral #1\relax}}

\begin{document}

\title{Quantum metrology in the presence of correlated noise via Markovian embedding}

\author{Arpan Das}
\email{arpan.das@uab.cat}
\affiliation{Faculty of Physics, University of Warsaw, Pasteura 5, 02-093 Warszawa, Poland}
\affiliation{F\'isica Te\`orica: Informaci\'o i Fen\`omens Qu\`antics, Department de F\'isica, Universitat Aut\`onoma de Barcelona, 08193 Bellaterra (Barcelona), Spain}

\author{Rafa{\l} Demkowicz-Dobrza{\'n}ski}
\email{demko@fuw.edu.pl}
\affiliation{Faculty of Physics, University of Warsaw, Pasteura 5, 02-093 Warszawa, Poland}

\begin{abstract}
We analyze quantum metrological protocols, where the sensing system is linearly coupled to a bosonic environment, by performing a Markovian embedding of the problem based on pseudomode formalism. This allows us to effectively model the problem using low-dimensional environment and apply recently developed powerful tools that yield optimal metrological protocols and fundamental metrological bounds for correlated-noise models. We illustrate the method by investigating a frequency estimation protocol in the presence of noise modeled effectively as a 
damped Jaynes-Cummings dynamics.
\end{abstract}

\maketitle

\section{Introduction}
One of the most prominent challenges, that the development of current quantum technologies face is environmental noise \cite{nielson, milburn2003, Suter2015, Preskill2018quantumcomputingin}. Accordingly, a fundamental question in quantum metrology \cite{Giovaennetti2006, Giovannetti2011, Demkowicz-Dobrzanski2015a, Toth2014, Degen2017, Braun2018, Pezze2018, Pirandola2018, Huang2024, Jiao2025, MONTENEGRO2025} is to determine the extent to which noise affects the estimation protocols equipped with full quantum sensing potential. 

In presence of \emph{uncorrelated} (Markovian) noise, there is a plethora of results \cite{Zhou2017, Fujiwara2008, Escher2011, Demkowicz-Dobrzanski2012, Koodynski2013, Demkowicz-Dobrzanski2014, Sekatski2016, Demkowicz-Dobrzanski2017, Zhou2019e, Zhou2020, adaptive-new-bound, Arpan2025}  that have provided comprehensive solutions to this question, including efficiently computable 
fundamental bounds on achievable precision as well as 
methods to find the optimal sensing protocols. 

In presence of \emph{correlated} noise, however, the problem is much more challenging. A number of papers have addressed this problem for particular models, restricting the generality of  metrological protocols considered, to make the solution feasible \cite{Chin2012, Szankowski2014, Macieszczak2015, Smirne2016, Beaudoin2018, Tamascelli2020, Riberi2022}. One of the most common simplification, 
is to consider non-Markovian interaction \cite{Vega2017, Milz2021, Shrikant2023} of the sensing probe with the environment over certain period of time, and compute the resulting extractable information from the probe. Based on this result, one may assess the information that would be accessible over the course of a long experiment, where the probe will be exposed to interactions with the environment and measured multiple-times. This approach, however, makes an assumption that in each of the sensing-steps the probe interacts with a freshly-prepared environment. Apart from the so-called collisional models \cite{Ciccarello2022}, this is hardly justified in real non-Markovian environments, where typically the probe will interact with the same environment, and hence environmental induced correlations between different sensing steps will be present in general.
Moreover, if one attempts to tackle the problem in full generality, one should consider the most general adaptive metrological protocols, where in fact the inter-step environmental correlation may be crucial to design the optimal protocol. If we neglect these aspects, we are following a simplified framework, which we will refer to as the \emph{fresh-environment} approach, in this paper.

Recently, there has been a surge of interest in investigating this correlated-noise problem in quantum metrology problem with utmost generality, without being confined to specific case studies \cite{Yang_2019, Altherr_2021, Liu2023,  Kurdzialek_2024, Liu2024efficienttensor, kurdzialek-2025universalboundsquantummetrology}. The approach is based on the mathematical framework of quantum combs \cite{Chiribella2008b, Chiribella2009}, which describes the most general quantum adaptive strategies as well as the parameter encoding, encompassing any types of correlated or noncorrelated-noise models. 

More concretely, given some parameter dependent dynamics of the sensing probes, the goal is to find the optimal preparation, quantum controls, measurements that result in maximization of acquired parameter information, typically quantified in terms of the Quantum Fisher Information (QFI) \cite{Helstrom1976, Holevo}.
The optimization task is extremely difficult especially in the limit of large number of probes or long sensing times, and as a result, large number of quantum control operations (or both!). The technique of quantum comb was first adopted \cite{Yang_2019, Altherr_2021, Liu2023} to construct numerical methods for finding optimal adaptive strategies suited for small number of quantum controls involved, primarily based on Maximization over Purification (MOP) techniques \cite{Fujiwara2008, Demkowicz-Dobrzanski2012}. These techniques, however, are not efficient for large number of quantum controls applied. 

Recently, this approach has been reconciled with the Iterative see-saw (ISS) QFI optimization algorithm \cite{Macieszczak2013, Toth2018} and the tensor-network framework \cite{Kurdzialek_2024}. The inherent tensor network structure of this approach allows one to construct efficient numerical optimization protocols for large number of quantum controls and hence, optimize performance over long sensing times. While, the authors applied this techniques to analyze metrological models involving a simple correlated-dephasing noise, this was more of an illustration of the techniques, as the model did not represent a physically-relevant noise that might be encountered in real-life applications. Moreover, the character of correlations was purely classical and could not be interpreted as the representation of some truly quantum non-Markovian dynamics.
\begin{figure}[t]
\includegraphics[width=\columnwidth]{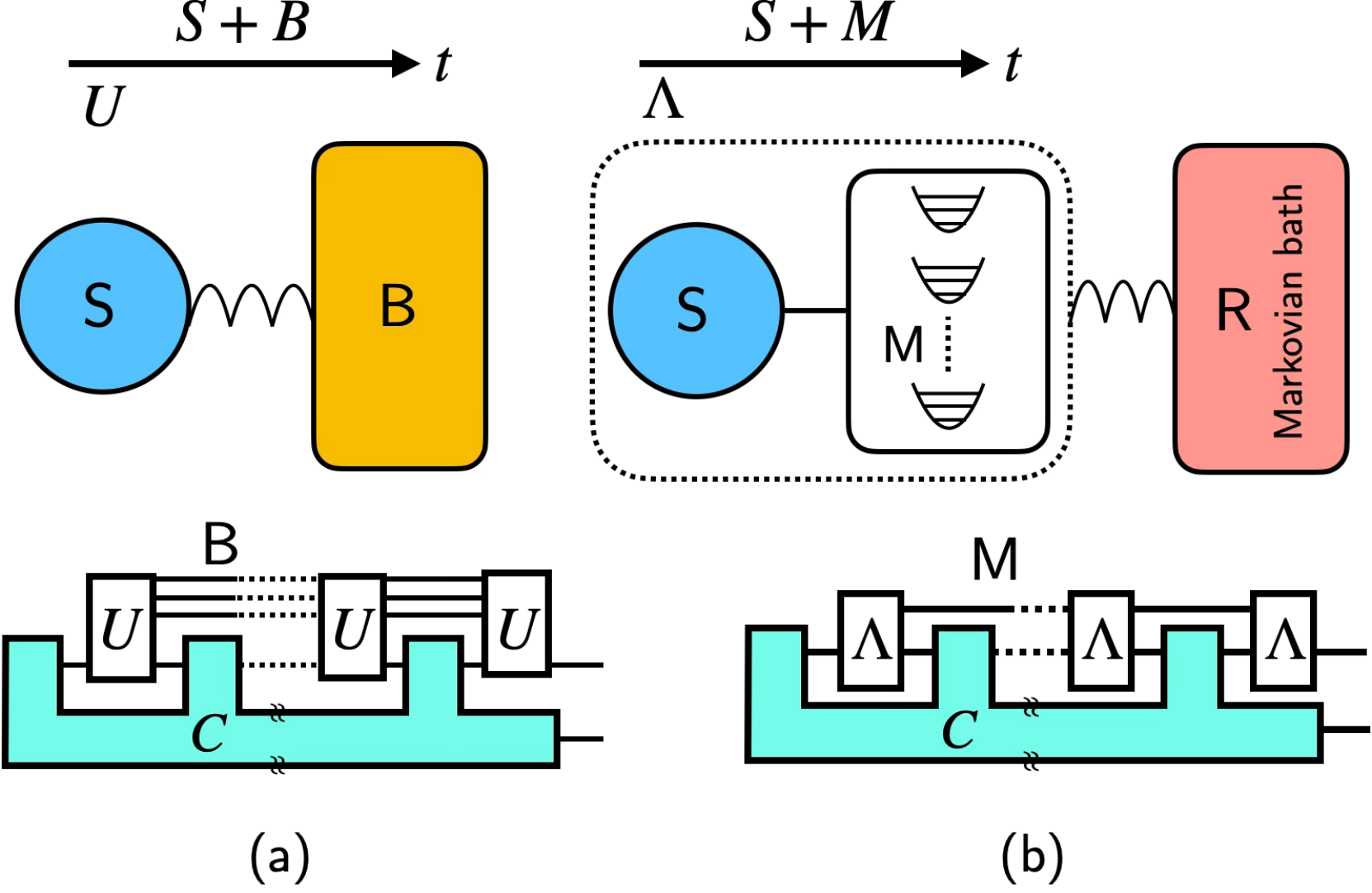}
    \caption{Schematic  illustration of two equivalent setup leading to same reduced dynamics of the system. (a) System is linearly coupled to a bath (infinite) with arbitrary coupling strength. The total system-bath dynamics is governed by the Unitary $U$, generated by the total Hamiltonian $H$ as in Eq. (\ref{original_H}). The corresponding metrological protocol (below) should in principle involve the whole bath-environment (inaccessible) to account for the correlation built up in course of the dynamics. (b) System space enlarged with a finite number of discrete pseudomodes together obeys a Markovian master equation as in Eq. (\ref{sm-master}) ($\Lambda$ is the corresponding dynamical map)  such that the reduced system dynamics is the same as the original one. The metrological protocol (below) now involves finite pseudomode space $M$ (inaccessible) to effectively model all the correlations present in the model.}
\label{fig:intro}
\end{figure}

In this paper, we show a way to apply the tools developed in
\cite{Kurdzialek_2024, kurdzialek-2025universalboundsquantummetrology}
to analyze quantum metrological scenarios involving 
physically-relevant and inherently quantum correlated noise models.
Our method is based on a particular technique of Markovian embedding. We start with the dynamics of a general open quantum system linearly coupled to a bosonic bath, where the parameter to be estimated is encoded in the Hamiltonian of the system. In general, the reduced dynamics of the open system \cite{breuer02}  is highly non-Markovian and in most cases an exact solution does not exist as it involves dealing with infinite number of degrees of freedom. Consequently, any metrological task that involves the reduced system dynamics, is not possible unless some standard approximations are to made to be within the Markovian regime. This potential difficulty can somehow be circumvented  using the fact that the system dynamics is fully determined by the bath correlation function (assuming a Gaussian initial state of the bath), which allows one to effectively describe the system dynamics through the Markovian dynamics of an enlarged system. In this formalism, the original environment $B$ is replaced by a finite number of virtual modes $M$ (called pseudomodes) \cite{Garraway1997, Dalton2001, Tamascelli2018, Pleasance2020, Lambert2019, pleasance2021pseudomode}, which together with the system forms the enlarged system following a Lindblad master equation, such that the reduced dynamics of the system density matrix is identical to the original dynamics, see Fig. \ref{fig:intro}.
This method may also be extended to include fermionic bath \cite{Chen_2019}, but in this paper we only focus on bosonic bath. The advantage of this formalism is that, now we have a master equation for the dynamics where the infinite bath has been effectively replaced by a finite number of degrees of freedom (number depends on the bath correlation function). This allows one to deal the quantum correlated noise in a less resource intensive and apply the quantum comb formalism both for QFI optimization and derivation of fundamental bounds. We can also study how the presence of noise correlation between different steps of a protocol affects the achievable QFI, by comparing the results with the \textit{fresh-environment} simplified approach, we have mentioned above. 

%After discussing the general framework of ISS scheme in combination with quantum comb for arbitrary open quantum system, we exemplify our approach for damped Jaynes-Cummings model. %where an exact master equation exists. 

This paper is organized as follows. In section \ref{section-2} we discuss the preliminaries regarding QFI optimization involving quantum comb techniques. In Sec. \ref{section-3}, we use the pseudomode model in conjunction with quantum comb for QFI optimization. In section \ref{section-4}, we exemplify our formalism for damped Jaynes-Cummings model. Finally, we conclude in section \ref{section-5}.

\section{Quantum comb formalism for optimal adaptive protocols}
\label{section-2}
A paradigmatic theoretical quantum metrological task is typically formulated as estimating a parameter $\varphi$, encoded in a quantum channel $\Lambda_\varphi: \mathcal{L}(\mathcal{H}_{\rm in})\rightarrow \mathcal{L}(\mathcal{H}_{\rm out})$, where $\mathcal{H}_{\rm in}$, $\mathcal{H}_{\rm out}$ are the Hilbert spaces for input and output quantum systems respectively and $\mathcal{L}(\mathcal{H})$ is the set of linear operators on the Hilbert space $\mathcal{H}$. The channel acts on a $\varphi$-independent initial probe state $\rho_0\in\mathcal{L}(\mathcal{H}_{\rm in}\otimes \mathcal{H}_{\rm anc})$ to produce an output state $\rho_\varphi\doteq(\Lambda_\varphi\otimes \mathds{1})\rho_0\in \mathcal{L}(\mathcal{H}_{\rm out}\otimes \mathcal{H}_{\rm anc})$, which is then measured with a generalized measurement $\{M_i\}$, such that the outcome $i$ occurs with probability $p_\varphi(i)={\rm Tr}(\rho_\varphi M_i)$. Here, $\mathcal{H}_{\rm anc}$ denotes the Hilbert space of ancillary system. The parameter $\varphi$ is estimated with an estimator $\tilde{\varphi}(i)$, with the goal to find protocols which give results closest to the true value of $\varphi$. For a locally unbiased estimator $\tilde{\varphi}(i)$, we have the lower bound on achievable estimation variance, known as the quantum Cra\'{m}er-Rao bound \cite{Helstrom1976,Braunstein1994},
\begin{equation}
\Delta^2 \tilde{\varphi}\geq \frac{1}{F_Q(\rho_\varphi)},
\end{equation}
where, $F_Q(\rho_\varphi)$ is the quantum Fisher information (QFI). 

Consequently, the problem of finding optimal estimation protocols reduces to the maximization of output state QFI: $F_Q(\Lambda_\varphi)={\rm max}_{\rho_0}F_Q(\rho_\phi)$, referred to as the \emph{channel QFI}. When multiple channels are probed simultaneously, the full potential of quantum metrology can be exploited using arbitrary quantum controls in between the sequential use of quantum channels \cite{Giovaennetti2006, Demkowicz-Dobrzanski2014, Zhou2020, adaptive-new-bound}. This strategy, known as adaptive or active quantum feedback (AD) is the most general strategy encompassing all other. Formally, a general AD strategy (along with the input probe state) for $N$ channels can be represented as a single Choi-Jamio{\l}kowski (CJ) operator \cite{Bengtsson2006} $C^{(N)}\equiv C$, known as quantum comb \cite{Chiribella2008b, Chiribella2009},
\begin{equation}
    C\in{\rm comb}[(\varnothing, \mathcal{H}_1), (\mathcal{H}_2, \mathcal{H}_3), \dots, (\mathcal{H}_{2N-2}, \mathcal{H}_{2N-1} \otimes \mathcal{A}_N],
\end{equation}
where, each pair of Hilbert spaces represents the respective input (labeled by even number) and output (labeled by odd number) Hilbert spaces of each tooth of the comb with $\varnothing$ denoting no input and $\mathcal{A}_N$ representing the ancilla space after {$N$th} use of the channel. The normalization conditions for $C$ are,
\begin{align}
    & C \succeq 0,~~ C^{(0)}=1,\\
    & {\rm Tr}_{2k-1}[C^{(k)}]=\mathds{1}_{2k-2}\otimes C^{(k-1)},~~~k\in \{1,\dots,N\}.
\end{align}
\begin{figure}[t]
\includegraphics[width=\columnwidth]{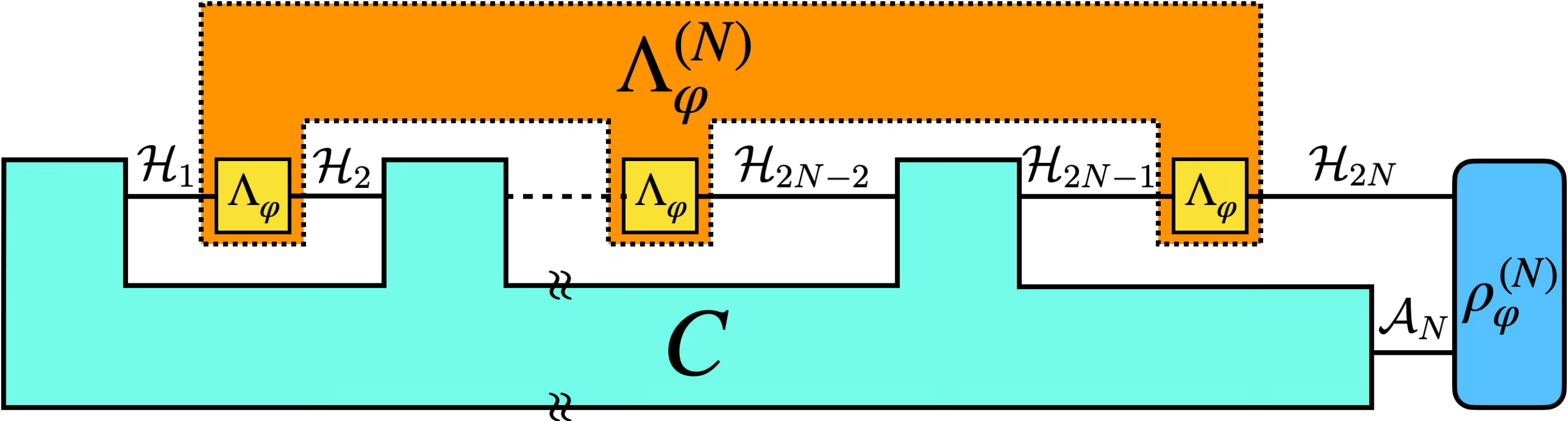}
    \caption{Schematic of general adaptive metrological protocol involving $N$ independent ($\Lambda_\varphi^{\otimes N}$) or correlated quantum channels ($\Lambda_\varphi^{(N)}$) with arbitrary quantum controls ($C$).}
\label{fig:intro-2}
\end{figure}

These conditions encode the causal structure of the comb, guaranteeing that concatenation of the respective output and input spaces with quantum channel will yield a physical quantum channel as well. Analogously, the action of $N$ sensing channels may also be expressed as a quantum comb $\Lambda_\varphi^{(N)}\in {\rm comb}[(\mathcal{H}_1, \mathcal{H}_2),\dots, (\mathcal{H}_{2N-1}, \mathcal{H}_{2N})]$, which in general represents any type of noise and signal correlation, where for uncorrelated noise, $\Lambda_\varphi^{(N)}=\Lambda_\varphi^{\otimes N}$. Finally, combining the parameter encoding and the controls, the output state is $\rho_\varphi^{(N)}=C\star\Lambda_\varphi^{(N)}\in \mathcal{\mathcal{H}}_{2N}\otimes \mathcal{A}_N$, where $\star$ denotes the link product, formally representing the concatenation of respective quantum combs, see Fig.~\ref{fig:intro-2}. 

The optimal estimation protocol should 
maximize the QFI for the output state $\rho_\varphi^{(N)}$ over controls $C$: $F_{\rm AD}^N=\max_{C} F(C\star\Lambda_\varphi^{(N)})$. There are two different approaches to carry out this task: the Minimization Over Purifications approach (MOP) and the Iterative See-Saw (ISS approach) \cite{Kurdzialek_2024}. 
The MOP approach \cite{Yang_2019, Altherr_2021, Liu2023} allows to identify the optimal adaptive protocol via a single semi-definite programme (SDP), but is limited to small 
dimensional problems with few quantum controls. 
On the other hand, the ISS approach \cite{Kurdzialek_2024, Liu2024efficienttensor}, that we will pursue in this paper,  
allows for an efficient optimization of adaptive strategies involving multiple quantum control steps, thanks to the use of tensor-network optimization framework. Although, the optimality of the solution found needs to be double-checked either by performing multiple-numerical optimization runs, or via comparison with the fundamental bounds. Apart from the ISS optimization, we will 
also use recent results to derive fundamental bounds in presence of correlated noise \cite{kurdzialek-2025universalboundsquantummetrology}, that in fact, are obtained by a creative development of the MOP ideas. In case of a single-channel QFI ISS optimization  we first define the pre-QFI function,
\begin{equation}
    F(\rho_0,L)=2{\rm Tr}\left(\dot{\rho}_\varphi L\right)-{\rm Tr}\left(\rho_\varphi L^2\right)
\end{equation}
Maximization of $F(\rho_0,L)$, over the input state $\rho_0$ and Hermitian $L$ results in the channel QFI \cite{Macieszczak2013},
\begin{equation}
    F_Q(\Lambda_\varphi)=\max_{\rho_0, L}F(\rho_0, L).
\end{equation}
This double maximization can be evaluated efficiently in two iterative steps. We start with a random initial state for which $F(\rho, L)$ is maximized over $L$ to obtain optimal $L_0$. In the next step, we fix $L$ to be $L_0$ and maximize $F(\rho, L_0)$ over $\rho$ to obtain optimal $\rho_0$. These two iterative steps are repeated until the pre-QFI function converges, which can be justified for generic input states. Both of the steps can be formulated as simple SDP's. 

When searching for the optimal adaptive strategy involving  $N$ uses of channels, using the 
ISS optimization \cite{Kurdzialek_2024, Liu2024efficienttensor}, we define the pre-QFI function as,
\begin{equation}
    F^{(N)}(C,L)=2{\rm Tr}\left(\dot{\rho}_\varphi^{(N)} L\right)-{\rm Tr}\left(\rho_\varphi^{(N)} L^2\right),
\end{equation}
where, as before, $\rho_\varphi^{(N)}=C\star\Lambda_\varphi^{(N)}$. The optimal QFI is obtained as a double maximization of this pre-QFI function,
\begin{equation}
    F_Q(\Lambda_\varphi^{(N)})=\max_{C, L}F^{(N)}(C, L),
\end{equation}
which again can be posed as a SDP. In this form though, the efficiency of optimization is comparable to MOP approach as the computation complexity grows exponentially with $N$. The real advantage of ISS is that it can be adapted to tensor network framework and the dimensions $d_A$ of ancillary systems are controllable. One can split the whole control $C$ into smaller teeth $C_i$'s, such that,
\begin{equation}
    C=C_1\star C_2\star\cdots C_N,
\end{equation}
where, $C_1\in \mathcal{L}(\mathcal{H}_1\otimes \mathcal{A}_1)$ and $C_k\in \mathcal{L}(\mathcal{A}_{k-1}\otimes \mathcal{H}_2\otimes \mathcal{H}_3\otimes \mathcal{A}_N)$ for $2\leq k\leq N$. In principle to realize all possible $C$, the dimensions of ancillas required, grow exponentially with $N$. However, in practice it is often enough to consider fixed and small $d_A$ to get the optimal (or close to optimal) strategy. This makes the ISS procedure well suited to analyze the problems in the large $N$ limit. The correlated channel $\Lambda^{(N)}$ can also be decomposed in a similar way, with $\mathcal{E}_k$ denoting the environmental space (in place of $\mathcal{A}_k)$ to reproduce the correlated noise. When noise is not correlated, the environmental links can be ignored as $\Lambda_\varphi^{(N)}=\Lambda_\varphi^{\otimes N}$. Both classical and quantum correlated noise can be dealt in this way to find optimal metrological protocols. However, so far only classically correlated noise has been studied in using this approach \cite{Kurdzialek_2024}. In the next section, we show how to apply the Markovian embedding ideas in order to map complex quantum metrological models involving quantum correlated noise, into model that can be efficiently analyzed using the tensor-network approach introduced above.

\section{Markovian embedding using pseudomode theory in a quantum  metrological correlated noise scenario}
\label{section-3}
Our goal here is to device a general framework such that the tensor network formalism for ISS procedure can be applied to an important class of quantum metrological models, where the system is linearly coupled to a bosonic environment.
Consider a problem of estimating a parameter 
$\Omega$ of a system Hamiltonian $H_S(\Omega)$, where the total Hamiltonian $H$ represents the dynamics of a system, linearly coupled with a bath as, 
\begin{equation}
\label{original_H}
    H=H_S(\Omega)+H_B+H_{SB},
\end{equation}
%In the above, $H_S(\varphi)$ is the system (that we control) Hamiltonian with encoded parameter $\varphi$ to be estimated, 
where $H_B$ is the bath Hamiltonian and $H_{SB}$ is the interaction Hamiltonian between the system and the bath. Note, that we use $\Omega$ instead of $\varphi$ we used before to parametrized channels $\Lambda_\varphi$, to indicate that typically we deal with a frequency $\Omega$ estimation problem, which only can be translated into a phase $\varphi$ estimation problem after we integrate the dynamics over some specific time $t$, and identify $\varphi=\Omega t$.
%In what follows, we will typically drop the explicit dependence on $\Omega$ for notational convenience. 

For the present study, we restrict ourselves to bosonic bath with $H_B=\sum_k \omega_k a_k^\dagger a_k$, but the analysis can also be extended to a fermionic bath. Here, $a_k$ and $a_k^\dagger$ are the corresponding annihilation and creation operator, satisfying $[a_i, a_{j}^\dagger]=\delta_{ij}$. The interaction Hamiltonian is of the form $H_{SB}=S\otimes B$, where $S=S^\dagger$ is the system operator and $B=\sum_k g_k(a_k+a_k^\dagger)$ is the bath operator. The initial state of the total system-bath setup is a factorized state of the system and the bath: $\rho_{SB}(0)=\rho_S(0)\otimes \rho_B(0)$. Furthurmore, the initial environment state $\rho_B(0)$ is assumed to be both Gaussian and stationary, meaning $[H_B, \rho_B(0)]=0$. Due to this, the reduced system dynamics,
\begin{equation}
\label{ori-dyn}
    \rho_S(t)={\rm Tr}\left(e^{-iH t}\rho_S(0)\otimes \rho_B(0)e^{iH_t}\right)
\end{equation}
is completely determined by the two-time correlation function (assuming ${\rm Tr}(\rho_B(0)B)=0$) \cite{Tamascelli2018},
\begin{equation}
\label{eq:ct}
    C(\tau)=\langle B(\tau) B(0)\rangle={\rm Tr}(B(\tau)B\rho_B(0)),
\end{equation}
where, $B(t)=e^{iH_B t}B e^{-iH_B t}$. Without loss of generality, we take $\rho_B(0)=\ket{0}\bra{0}_B$, which is the ground state of the environment, ensuring ${\rm Tr}(\rho_B(0)B)=0$.
With $\gamma(\omega)$ as the Fourier transform of the correlation function $C(\tau)$, we can write $C(\tau)$ as the inverse Fourier transform,
\begin{equation}
\label{ori-corr}
    C(\tau)=\frac{1}{2\pi}\int_{-\infty}^{\infty}\gamma(\omega)e^{-i\omega t}d\omega.
\end{equation}
$\gamma(\omega)$ is also known as the bath spectral density. Under the assumption that $\gamma(\omega)$ is a meromorphic function when analytically continued to the lower half complex $\omega$ plane and decays faster than $O(1/|\omega|)$ when $|\omega|\rightarrow \infty$ \cite{Pleasance2020, pleasance2021pseudomode}, $C(\tau)$ can be evaluated analytically in terms of the poles and residues of $\gamma(\omega)$. Applying the residue theorem, one can show that,
\begin{equation}
\label{correlation-exponential}
    C(\tau)=-i\sum_{l} r_l e^{-iz_l\tau},
\end{equation}
where, $z_l=\xi_l-i\lambda_l$ are the poles of $\gamma(\omega)$, and $r_l$'s are the corresponding residues. While the evaluation of the reduced system dynamics seems pretty straightforward, in practice, exact solutions (or a master equation description) are almost impossible (except some special cases) due to the presence of infinite number of degrees of freedom. One approach to deal with this difficulty is to map this highly non-Markovian dynamics to a simpler Markovian (hence we have a master equation) one in an enlarged system space. For the setup of Eq. (\ref{original_H}), we now describe the alternative model leading to exactly the same dynamics of $\rho_S(t)$. 

In this setup, the original environment is replaced with a finite number of discrete bosonic modes, which are known as pseudomodes, determined by the poles and residues of $\gamma(\omega)$. The original system space ($S$) is now enlarged by these pseudomodes ($M$) with a total Hamiltonian \cite{Pleasance2020, pleasance2021pseudomode},
\begin{equation}
    H^\prime=H_S(\Omega)+H_M+H_{SM},
\end{equation}
where, $H_S$ is the system Hamiltonian as before, $H_M=\sum_l \xi_l b_l^\dagger b_l$ is the pseudomodes Hamiltonian, and $H_{SM}=S\otimes \sum_l \sqrt{-ir_l}(b_l+b_l^\dagger)$ is the interaction Hamiltonian between the system and the pseudomodes. The bosonic annihilation (creation) operator $b_l$ ($b_l^\dagger$) of $l$-th psedomode satisfy $[b_k,b_l]=\delta_{kl}$ as usual. For our purpose it is enough to assume that $\sqrt{-ir_l}$ is real, such that $H'$ is Hermitian. But the procedure can also be extended to include complex coupling of the system with the pseudomodes \cite{pleasance2021pseudomode}.
Now, the pseudomodes are individually coupled to a Markovian bath ($R$) (see Fig. \ref{fig:intro}), such that the Hamiltonian for the extended model is given as \cite{pleasance2021pseudomode},
\begin{equation}
    H_{SMR}=H'+H_R+H_{MR},
\end{equation}
where,
\begin{align}
    & H_R=\sum_l \int \omega a^{\dagger}_{Rl}(\omega)a_{Rl}(\omega)d\omega\\
    &H_{MR}=\sum_l\int\sqrt{\frac{\lambda_l}{\pi}}\left(b_l^\dagger a_{Rl}(\omega)+b_l a^\dagger_{Rl}(\omega)\right)d\omega.
\end{align}
In the above, $a_{Rl}(\omega)$ ($a^\dagger_{Rl}(\omega)$) is the annihilation (creation) operator for frequency $\omega$ in the bath $R$ for the corresponding pseudomode $l$, satisfying $[a_{Rl}(\omega), a^\dagger_{Rl'}(\omega')]=\delta_{ll'}\delta(\omega-\omega')$. Density matrix $\rho_{SMR}(t)$ in the extended Hilbert space follows the unitary evolution,
\begin{equation}
    \rho_{SMR}(t)=e^{-H_{SMR}t}\left(\rho_{SM}(0)\otimes\ket{0}\bra{0}_R\right)e^{H_{SMR}t}
\end{equation}
One can now show that \cite{pleasance2021pseudomode} the reduced dynamics of $\rho_{SM}(t)={\rm Tr}_R \rho_{SMR}(t)$ follows the following Markovian master equation,
\begin{equation}
\label{sm-master}
    \frac{d}{dt}\rho_{SM}(t)=\mathcal{L}[\rho_{SM}(t)]=-i[H^\prime,\rho_{SM}(t)]+\mathcal{L}_D[\rho_{SM}(t)]
\end{equation}
where the disspative superoperator is given as,
\begin{equation}
    \mathcal{L}_D[\rho_{SM}]=2\sum_l\lambda_l\left(b_l\rho_{SM}b_l^\dagger-\frac{1}{2}\{b_l^\dagger b_l,\rho_{SM}\}\right).
\end{equation}
The reduced system dynamics for this master equation, obtained from a factorized initial state $\rho_{SM}(0)=\rho_S(0)\otimes \ket{0}\bra{0}_M$ is given as,
\begin{equation}
\label{aux-sys}
    \rho'_S(t)={\rm Tr}_M\left(e^{\mathcal{L}t}\rho_S(0)\otimes \ket{0}\bra{0}_M\right),
\end{equation}
where $\ket{0}\bra{0}_M$ is the ground state of the pseudomode Hamiltonian. Furthermore, initially, if the Markovian bath ($R$) is also in the ground state (and factorized state with $M$ and $S$), the above dynamics of the system density matrix will be solely determined by the correlation function of the joint $MR$ system (note that $SMR$ dynamics is unitary), which we denote as $C'(\tau)$. 
It can be shown that \cite{Pleasance2020, pleasance2021pseudomode} $C'(\tau)$ is identical to the correlation function $C(\tau)$ (Eq. (\ref{ori-corr})) of the original dynamics, which establishes the exact equivalence between the reduced dynamics obtained from Eq. (\ref{aux-sys}) and the original reduced dynamics of the system (Eq. (\ref{ori-dyn})), i.e. $\rho_S(t)=\rho'_S(t)$.  Of-course there are cases (like for thermal baths) where correlation functions $C(\tau)$ cannot be written as a finite sum of complex exponentials like in Eq. (\ref{correlation-exponential}), which means that $C(\tau)$ can not be exactly matched to $C'(\tau)$. In those cases, one optimizes the parameters to get $C'(\tau)$ as close as possible to $C(\tau)$ \cite{Mascherpa2017, Tamascelli2018, Trivedi2021} such that $\rho_S(t)\approx \rho'_S(t)$. Nevertheless, in the following section we discuss an example where exact equivalence between $\rho_S(t)$ and $\rho'_S(t)$ is obtained. The advantage of this method is that now we have a master equation which involves finite number of modes instead of an infinite environment, greatly simplifying numerical computation. For example for a spectral density $\gamma(\omega)$, which has only one pole, we have one pseudomode, which makes the master equation just a two-qubit evolution for a qubit system. 

If we integrate the joint dynamics of SM system over time $t$ we may, therefore, obtain an effective parameter dependent channel 
\begin{equation}
    \Lambda_{\Omega,\Delta t} = e^{\mathcal{L} \Delta t},
\end{equation}
where $\mathcal{L}$ is the Liouvillian superoperator appearing in the effective master equation \eqref{sm-master}.
Provided the effective dimensions of systems $S$ and $M$ are small enough, we may apply the ISS tensor-network optimization methods discussed in Sec.~\ref{section-2}. We may consider, evolution of a system over a long time $t=N \Delta t$, where general quantum controls on the system $S$ (and ancillary systems $A$) are allowed every time interval $\Delta t$. In this way, the dynamics of the SM system will be described in terms of a quantum comb $\Lambda_{\Omega,\Delta t}^{(N)}$, for which we may try to find the comb $C$ representing the optimal adaptive protocol. The pseudomode space $M$ works as the inaccessible memory (or the environment), but will play an important role as a mediator of correlations between one quantum control operation to another. As a result, it will affect both the form as well as the performance of the  optimal  metrological protocols. 
Demonstrating the role of this environment mediated correlations in the performance of the metrological protocol is the main purpose of this paper, which is presented in the following section. 
\section{Two-level atom frequency estimation for a  damped Jaynes-Cummings model}
\label{section-4}
In this section, we study an example which has an exact analytical solution. The model we consider is spontaneous decay of two-level system in a bosonic bath at zero temperature. The total Hamiltonian of the setup is given by,
\begin{equation}
\label{total-hamiltonian}
    H=H_S(\Omega)+H_B+H_{SB},
\end{equation}
where,
\begin{align}
    \label{hamiltonians}
    & H_S(\Omega)=\frac{\tilde{\omega}\Omega}{2} \sigma_z\otimes \mathds{1},\\
    &H_B= \mathds{1}\otimes\sum_k \omega_k b_k^\dagger b_k,\\
    \label{interaction-H}
    &H_{SB}=\sum_k \left(g_k(\sigma_+ \otimes b_k)+ g_k^*(\sigma_-\otimes b_k^\dagger)\right).
\end{align}
Without loss of generality, we set $\tilde{\omega}=1$, which basically sets the natural timescale of the problem to be $\tilde{\omega}^{-1}$. The parameter $\Omega$ to be estimated represents the transition frequency between the two levels of system $S$ and we assume it to be dimensionless. As a result the QFI we compute will also be dimensionless.
Note that total Hamiltonian $H$ commutes with the total excitation number operator $N$, given by,
\begin{equation}
    \label{number-operator}
    N=\sigma_+\sigma_-\otimes \mathds{1}+\mathds{1}\otimes \sum_k b_k^\dagger b_k,~~~~[H,N]=0.
\end{equation}
This means that total number of excitations is a constant of motion owing to the excitation preserving interaction Hamiltonian. This is an important requirement for an exact solution to exist \cite{breuer02}. 

Following the general prescription outlined in the previous section, 
%Our focus is to describe the non-Markovian dynamics of the system interacting with bath that are Gaussian and stationary (as mentioned in the previous section) with the method of pseudomodes. Specifically, 
we assume that $\rho_B(0)= \ket{0}\bra{0}_B$, where $\ket{0}_B$ is the vacuum state of the bath that satisfies the stationary condition, 
    $[H_B, \rho_B(0)]=0$.
Now, the reduced system dynamics is given as,
\begin{equation}
   \rho_S(t)= {\rm Tr}_B\left\{e^{-iHt}(\rho_S(0)\otimes \ket{0}\bra{0}_B)e^{iHt}\right\}.
\end{equation}
With ${\rm Tr}_B\{\rho_B(0)B(t)\}=\langle B(t)\rangle=0$, where $B(t)=e^{iH_B t}(\sum_k g_kb_k)e^{-iH_Bt}$, the reduced system dynamics is fully determined by the two point correlation function $C(\tau)$ or its Fourier transform 
$\gamma(\omega)$ as defined respectively in  \eqref{eq:ct} and \eqref{ori-corr}.
%,
%\begin{equation}
%    C(\tau)\equiv\langle B(t)B(t-\tau)\rangle= \frac{1}{2\pi}\int_{-\infty}^{\infty}\gamma(\omega) e^{-i\omega \tau}d\omega
%\end{equation}
%determined by the specific form of 
%$\gamma(\omega)$ function, see Eq.~\eqref{ori-corr}. %Note that, this method can be applied with any interaction Hamiltonian without the nedd that it is excitation preserving. But in this example, we will stick with the excitation preserving interaction Hamiltonian as in Eq. (\ref{interaction-H}). 
%Now we can proceed as before to describe the dynamics of the enlarged system and the psedomodes.
We choose the bath spectral function of a detuned Jaynes-Cummings model (two level system in a single cavity mode coupled to a bosonic environment) \cite{breuer02},
\begin{equation}
    \gamma(\omega)=\frac{\gamma_0\lambda^2}{(\omega_0-\omega)^2+\lambda^2}.
\end{equation}
Here, $\lambda$ is the spectral width of the coupling, which is related to the bath correlation time $\tau_B$ as 
\begin{equation}
\tau_B\sim \lambda^{-1},
\end{equation}
$\gamma_0$ is the system-bath coupling strength and the amount of detuning from the system frequency is $\Omega-\omega_0$.
For this choice of $\gamma(\omega)$, we have just one pole at the lower half of the complex $\omega$-plane,
\begin{equation}
    z_1=\omega_0-i\lambda, \quad  r_1 = i\frac{\gamma_0\lambda}{2}
\end{equation}
Hence, $C(\tau)=\frac{\gamma_0\lambda}{2}e^{-\lambda \tau}$. Now, according to the previous section, the system-pseudomode enlarged description is given as following \cite{Garraway1997, Tamascelli2018, Pleasance2020, pleasance2021pseudomode},
\begin{equation}
    H^\prime=H_S(\Omega) +H_M + H_{SM},
\end{equation}
where,
\begin{align}
    & H_M=\mathds{1} \otimes \omega_0 b_1^\dagger b_1,\\
    & H_{SM}=\sqrt{\frac{\gamma_0\lambda}{2}}(\sigma_+\otimes b_1+\sigma_-\otimes b_1^\dagger),
\end{align}
where $b_1$, and $b_1^\dagger$ are bosonic annihilation and creation operators  of a single pseudo-mode $M$. 
This enlarged $SM$ system is connected with the Markovian bath through $M$ system only and now obeys the Markovian master equation,
\begin{equation}
\label{eff-master-1}
    \dot{\rho}_{SM}=-i[H^\prime,\rho_{SM}]+2\lambda\left(b_1\rho_{SM} b_1^\dagger -\frac{1}{2}\{b_1^\dagger b_1,\rho_{SM}\}\right).
\end{equation}

In order to apply the ISS tensor-network optimization approach presented in Sec.~\ref{section-2}, so that we can efficiently optimize multiple-step quantum adaptive protocols, we need to make sure that the Hilbert space dimensions required to describe the dynamics of $SM$ system are small enough.  
Thanks to excitation preserving property of the dynamics, if   system $M$ is initially prepared in the ground state $\ket{0}_M$, this implies that 
the whole dynamics will be restricted to at most a single total excitation, and hence $M$ may be assumed to be effectively described using a two dimensional space.% $\{\ket{0}_M,\ket{1}_M\}$. 

However, in general, the control operations, applied on the system, in the most general adaptive metrological protocol, may increase the total excitation numbers. 
Hence, in general, one might consider 
a larger-dimensional space $M$, where the dimension chosen would correspond to the maximal number of additional excitation we want to take into account as potentially appearing due to the applications of control operations. This will not qualitatively, change the analysis, but will obviously increase the numerical complexity of the problem. Since the prime goal of this paper is to show how the method works, we will not aim here to push the numerical aspects of the problem to its limit. Hence, we will restrict ourselves to the simplest case where $M$ is modeled  as a two dimensional system and leave  the analysis of the more general scenario for the future work. We will show, that even with this  restrictive scenario, we will already be able to demonstrate the impact of  environmental correlations on the optimal metrological performance that may be achieved in the model.
Consequently, we can replace $b_1$ and $b_1^\dagger$ with $\sigma_-$, and $\sigma_+$ respectively and 
arrive at the effective master equation of the model:
\begin{align}
\label{eff-master-2}
\nonumber
    &\dot{\rho}_{SM}=\mathcal{L}[\rho_{SM}(t)]=-i[H^\prime,\rho_{SM}]\\
    &+2\lambda\left((\mathds{1}\otimes\sigma_-)\rho_{SM} (\mathds{1}\otimes\sigma_+) -\frac{1}{2}\{\mathds{1}\otimes\sigma_+\sigma_-,\rho_{SM}\}\right),
\end{align}
where,
\begin{equation}    H^\prime=\frac{\Omega}{2}\sigma_z\otimes\mathds{1}+\mathds{1}\otimes\omega_0\sigma_z+\sqrt{\frac{\gamma_0\lambda}{2}}(\sigma_+\otimes \sigma_- +\sigma_-\otimes \sigma_+)
\end{equation}
%The solution to this master equation is,
%\begin{equation}
%\label{SM-dynamical-map}
%    \rho_{SM}(t)=\Lambda_t^{SM}[\rho_{SM}(0)],~~~\text{where}~~~\Lambda_t=e^{\mathcal{L}t},
%\end{equation}
%Now, this model can also be solved exactly with standard technique. We will use this exact solution to compare it with the correlated one (that with the pseudomode) with inaccessible memory in the metrological tasks. With total one initial excitation in the system-bath setup, the solution of this model is given as 
If we integrate  the above master equation we obtain the corresponding map:
\begin{equation}
\label{eq:smmaster}
\Lambda^{SM}_{\Omega,t} = e^{\mathcal{L} t}.
\end{equation}
Given the initial state $\rho_{SM}(0)=\rho_S(0)\otimes \ket{0}\bra{0}_M$, with $\ket{0}_M$ as the ground state of $H_M$, and assuming no quantum controls are applied, one may obtain an analytical form of the reduced system evolution
(in the interaction picture) \cite{breuer02},
\begin{multline}
\label{S-dynamical-map}
\rho_{S}(t)= \Lambda_{\Omega,t}^S[\rho_S(0)] = \t{Tr}_M\left( \Lambda_{\Omega,t}^{SM}[\rho_{SM}(0)] \right)=\\ =
    \begin{pmatrix}
        |c_1(t)|^2 & c_0^* c_1(t)\\
        c_0 c_1^*(t) & 1-|c_1(t)|^2,
    \end{pmatrix}
\end{multline}
where,
\begin{equation}
\label{eq:solrho}
\rho_S(0)=
    \begin{pmatrix}
        |c_1(0)|^2 & c_0^* c_1(0)\\
        c_0 c_1^*(0) & 1-|c_1(0)|^2,
    \end{pmatrix}
\end{equation}
and
\begin{equation}
\begin{split}
\label{c1-exp}
    c_1(t)=c_1(0)&e^{-(\lambda-i(\Omega-\omega_0)) t/2} \\
    & \left[\cosh\left(\frac{dt}{2}\right)+\frac{\lambda-i(\Omega-\omega_0)}{d}\sinh\left(\frac{dt}{2}\right)\right],
\end{split}
\end{equation}
with $d=\sqrt{\lambda^2-2\gamma_0\lambda-2i\lambda(\Omega-\omega_0)-(\Omega-\omega_0)^2}$.
The above form of the state, will serve us as a reference, to assess the metrological potential of autonomous dynamics of the system, where no controls are applied.
\begin{figure*}[t]
\includegraphics[width=2.0\columnwidth]{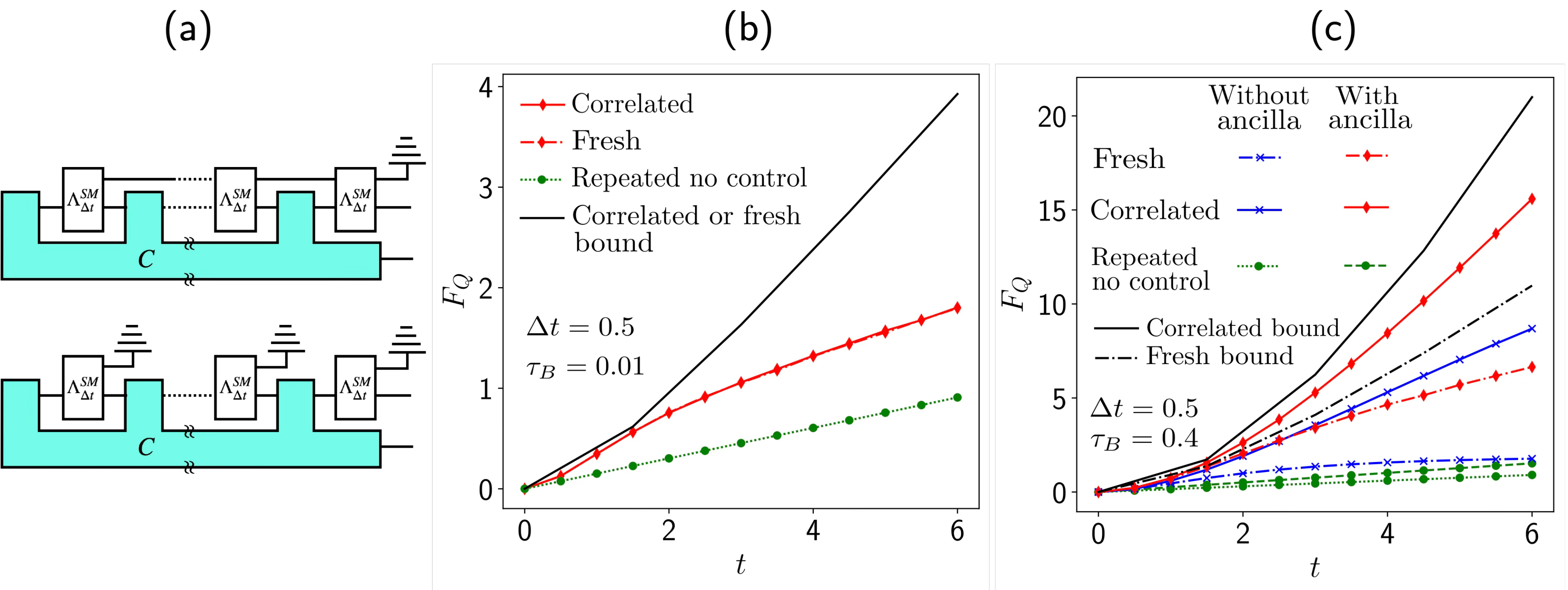}
    \caption{Comparison of QFI growth with time $t$ for three models, one with correlated environment (dynamics as given in Eq. (\ref{eq:smmaster})), one with fresh environment (dynamics as given in Eq. (\ref{S-dynamical-map})) and the third one is the no control scheme repeated over the course of time. (a) Schematic diagrams of the first two models. (b) QFI with qubit ancilla in the Markov limit; QFI with no control scheme repeated over the course of total sensing time and correlated bound for the dynamics with fresh/correlated environment (block sizes $=3$). Parameters for all the curves are $\gamma_0=2.457$, $\tau_B\sim\lambda^{-1}=0.01$ and $\Delta t=0.5$. (c) QFI for correlated, fresh and repeated no control schemes with qubit ancilla and without ancilla; bounds with correlated and fresh dynamics (block size $=3$). Parameter values for all the curves are $\gamma_0=2.457$, $\tau_B\sim\lambda^{-1}=0.4$ and $\Delta t=0.5$.}
    \label{fig:JC-model-plot}
\end{figure*}
For concreteness, we assume we are interested in local estimation of the parameter around $\Omega \approx 0$, and also take $\omega_0=0$ for simplicity. %This leads to $c_1(t)=c_1(0)e^{-\lambda t/2}\left[\cosh\left(\frac{dt}{2}\right)+\frac{\lambda}{d}\sinh\left(\frac{dt}{2}\right)\right]$ with $d=\sqrt{\lambda^2-2\gamma_0\lambda}\equiv d'$. 
With this choice, \eqref{eq:solrho} can also be regarded as the the solution of the following master equation described solely in terms of the systems $S$,
\begin{multline}
\label{master}
    \dot{\rho}_S(t)=-i\left[\frac{\Omega}{2}\sigma_z,\rho_S(t)\right]+\\+\gamma(t)\left(\sigma_-\rho_S(t)\sigma_+-\frac{1}{2}\left\{\sigma_+\sigma_-,\rho_S(t)\right\}\right)
\end{multline}
where,
\begin{equation}
    \gamma(t)=\frac{2\gamma_0\lambda\sinh(d t/2)}{d\cosh(d t/2)+\lambda\sinh(d t/2)}
\end{equation}
and 
\begin{equation}
d=\sqrt{\lambda^2-2\gamma_0\lambda}. 
\end{equation}
This is a spontaneous decay model of a two-level system, with a time dependent decay rate. 
It is important to note that, if we abstract out the parameter estimation problem ($\Omega$ dependence) for the moment, solely from the dynamical perspective, this master equation is identical to the one that can be obtained for $\rho_S(t)$ in Eq. (\ref{eq:solrho}) with $\omega_0=0$ and $\Omega=0$. Although, the QFI obtained for the dynamics of Eq. (\ref{master}) will not be same as the one obtained for the dynamics in Eq. (\ref{eq:solrho}) due to additional $\Omega$ dependence in the noisy part of the evolution. Nevertheless, strictly in the limit of $t\rightarrow 0$, this contribution will be negligible, and $\Omega$ dependent contribution will only come from the Hamiltonian part of the evolution same as in Eq. (\ref{master}). 
Now, in the weak coupling regime i.e., for $\gamma_0/\lambda<1/2$ (means bath correlation time is less than that of system's), $\gamma(t)$ is always positive and consequently the system dynamics is  Completely Positive divisible \cite{Rivas2014, Breuer2016, Li2018, CHRUSCINSKI2022}. This means that, 
the environmental correlations are not strong enough to cause the effective dynamics to appear as genuinely non-Markovian.  But, whenever $\gamma_0/\lambda>1/2$ (strong coupling regime), parameter $d$ is purely imaginary and $\gamma(t)$ is oscillatory with negative values indicating non-Markovian features. Also, one can see that for $t=\frac{2}{d'}\left[n\pi-\tan^{-1}\left(\frac{-id}{\lambda}\right)\right]$, we have diverging $\gamma(t)$, though the solutions remains always valid. This shows that the time-convolutionless master equation at those points does not exist. Now, in the Markov limit: $\lambda\gg \gamma_0$ (very short bath correlation time compared to system, i.e., no memory), $c_1(t)\approx c_1(0)e^{-\gamma_0 t/2}$, implying system dynamics reduces to the standard Markovian dynamics with constant decay rate. Finally, in the large time limit, the system goes back to its ground state.

In order to find the optimal strategy to estimate the parameter $\Omega$, encoded in the Hamiltonian, we employ an adaptive metrological protocol for multiple uses of the channels $\Lambda_{\Omega,\Delta t}^{SM}$, 
where we divide the total experiment time into $N$ steps, such that $t= N \Delta t$, see 
Fig. \ref{fig:JC-model-plot}(a). 
From one control operation to the next, the system-pseudomode setup follows the dynamics depicted by the dynamical map $\Lambda_{\Omega,\Delta t}^{SM}$ as in Eq. (\ref{eq:smmaster}). Note that, the smaller
$\Delta t$ we choose, in 
principle, the more general the strategy becomes, as the control operation are more frequent. We should expect, however, that arbitrary-fast controls will typically not be necessary, and depending on the model one may numerically find the time-scales on which controls provide real benefits. 
At the end of the protocol, we trace out the environment and perform possibly joint measurements to estimate the encoded parameter $\Omega$.

For the sake of understanding the role of environmental correlations, we also consider an alternative protocol, 
depicted in Fig. \ref{fig:JC-model-plot}(a) as the bottom scheme.
Here, after each step, we trace out environment $M$ and assume that at the subsequent step, environment $M$ is again prepared in a fresh input state $\ket{0}_M$. This means that in between two control operations, the system dynamics follow the dynamics described in Eq.~(\ref{S-dynamical-map}). This will allow us to examine the performance of the adaptive protocol without any memory effect caused by the system-environment correlation which is being carried forward. We call this a \emph{fresh-environment} model. Intuitively, it may be expected, that the model that includes environmental correlations, may offer 
possibility to extract more information on the system parameter, taking into account that part of the system information which went into the environment at certain time, may be partially recover at some later moments of the protocol.
We perform numerical optimization using ISS tensor-network  formalism  introduced in Ref.~\cite{Kurdzialek_2024}, and implemented recently in a publically available  QMetro++ package  \cite{Dulian2026qmetropython}. We numerically calculated QFI for multiple number of channels for both the correlated and fresh-environment cases. More conceretly,  for a given total sensing time $t$, we calculated the maximal achievable QFI for $N= t/\Delta t$ adaptive protocol steps, with some fixed ancillary space dimension.

Now, as a sanity-check, we first examine the Markov limit, which as discussed earlier is obtained in the limit $\lambda\gg\gamma_0$. In this limit the system dynamics described by $\Lambda_{\Omega,t}^S$ is basically a semigroup dynamics with a constant decay rate $\gamma_0$. This means that, there is absolutely no correlation between the system and the environment even in the case for joint dynamics for the system and the pseudomode. Consequently, in this limit both models should give same QFI, which we can indeed see in the Fig.~\ref{fig:JC-model-plot}(b).
Moreover, to show the effectiveness of the intermediate control operations, we consider the simplest scheme, where, in each step of duration $\Delta t$, the system evolves according to Eq. (\ref{S-dynamical-map}) and the corresponding optimal QFI is computed. The optimization can still involve ancillary system.
At each step, the protocol is renewed and is repeated for $N$ times with no control operations in between. Consequently, for the $N$ step protocol, the QFI we get is simply $N$ times the QFI for one single step. We refer this scheme as \textit{repeated no control} model which is also plotted in Fig.~\ref{fig:JC-model-plot}(b). As can be seen from the plot, this scheme produces much worse results compared to the schemes with control operations.
% Moreover, in this limit, our previous bound for Markovian quantum metrology \cite{Arpan-2025} is expected to hold when plotted for the amplitude damping dynamics with constant decay rate $\gamma_0$ as following,  
% \begin{equation}
% \label{master-markov}
%     \dot{\rho}_S(t)=-i\left[\frac{\Omega}{2}\sigma_z,\rho_S\right]+\gamma_0\left(\sigma_-\rho_S\sigma_+-\frac{1}{2}\left\{\sigma_+\sigma_-,\rho_S\right\}\right)
% \end{equation}
% While the validity of the bound is evident from the plot, it is not very tight.  

Additionally, we plot the fundamental correlated bound on QFI \cite{kurdzialek-2025universalboundsquantummetrology}, which is universal for any kind of correlated noise (including inherently quantum noise) 
and at least as tight (if not more) as the previous Markovian bound in the case of uncorrelated noise \cite{Arpan2025}. We plot this bound using elementary blocks involving three concatenated elementary channels for both correlated ($\Lambda_{\Omega, \Delta t}^{SM}$) and uncorrelated ($\Lambda_{\Omega,t}^{S}$) dynamics in the Markovian limit. As expected, they also overlap with each other and provide a valid upper bound for the QFI. Moreover, we will see later that when we are not in the Markovian limit, only the correlated bound provides a valid upper limit on the QFI. 
In Fig. \ref{fig:JC-model-plot}(c), we focus on the example, where environmental correlation are expected to become relevant. We choose $\Delta t =0.5$, and 
$\gamma_0 \approx \lambda = 2.5$, which implies that bath correlations $\tau_B \sim \lambda^{-1}= 0.4$ are comparable with the intervals between quantum control operations applied.  
We consider protocols, where no ancillary system is involved, and hence control operations are performed solely on the $S$ system, and when ancilla is a single qubit system, that may be entangled in a controlled way with $S$ over the course of the protocol. 

As anticipated, we see a big improvement in QFI, 
when environmental correlations are properly taken into account, and the advantage is most pronounced, when we employ an ancillary qubit in the protocol. In appendix \ref{appen-A}, we present a detailed description of the control operations for $N=2$ (without ancilla) and discuss how environmental correlations help to improve the QFI.
This highlights the crucial role of correlations in the model, which when not taken into account, leads to a huge underestimation of extractable parameter information. 
%Note Especially, for the fresh-environment case without ancilla, the QFI is almost close to zero. 
Like before, we plot the fundamental correlated bound using elementary blocks involving three concatenated elementary channels $\Lambda_{\Omega, \Delta t}^{SM}$,  which shows that the protocol involving a single qubit ancilla is close to the optimal performance. It is possible that the gap between the lower bound (the QFI we obtain from tensor network formalism in conjunction with ISS) and the upper bound might be further decreased, had we increased the ancilla dimension, or improve the tightness of the bound by increasing the block size used for computation of the bound  as described in \cite{kurdzialek-2025universalboundsquantummetrology}. We also plot the same bound using the elementary blocks of three concatenated channels $\Lambda_{\Omega,\Delta t}^S$. While this provides a valid upper bound to the QFI corresponding to the \textit{fresh-environment} model, QFI (calculated using qubit ancillary system) corresponding to the dynamics $\Lambda_{\Omega,\Delta t}^{SM}$ violates this. This shows the importance of the choosing a correct bound of QFI especially in the presence of correlation. 
\begin{figure}[t]
\includegraphics[width=\columnwidth]{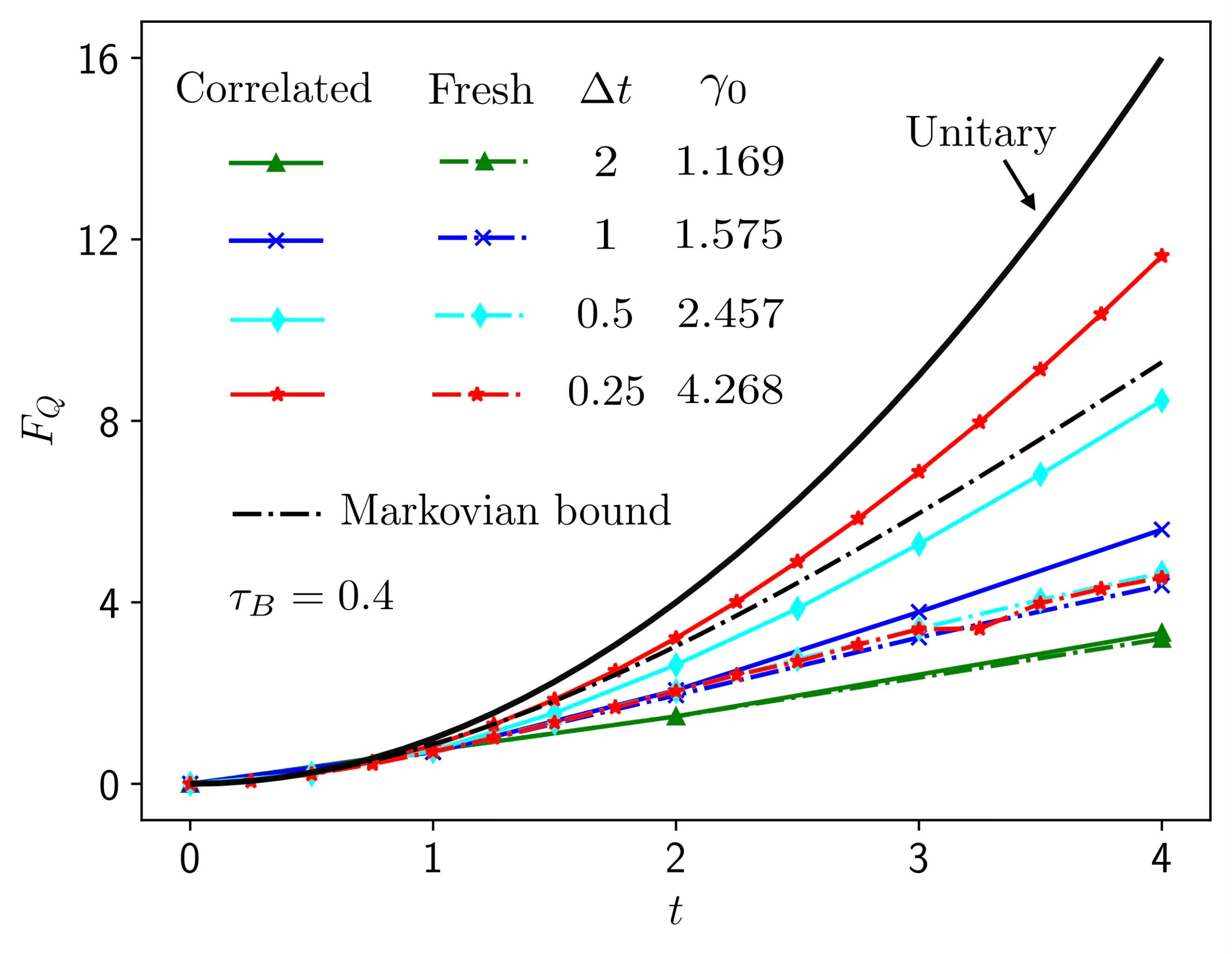}
    \caption{Plots for QFI with time $t$, for different $\Delta t$ and qubit ancilla, with fixed $\lambda=2.5$ ($\tau_B=0.4$) and different $\gamma_0$. Markovian bound (constant decay rate $\gamma_0 \approx 1.14$ ) % 1.14355 \rdd{too many digits}) 
    and the ultimate limit with unitary encoding are also illustrated in the figure.}
\label{fig:JC-model-plot-2}
\end{figure}
As noticed above, in order to observe the benefits of environmental correlations, we needed to operate in the 
regime, where bath correlations $\tau_B$ 
are long enough to be relevant on the time scale $\Delta t$, that separates subsequent applications of quantum controls. 
In order to investigate this aspect of the problem, we compare the correlated and the fresh-environment models, for varying values $\Delta t$, starting from $\Delta t=2$ which is significantly longer than the bath correlation time $\tau_B
\sim\lambda^{-1}= 0.4$, and reducing it to $\Delta t =0.25$ which close to half of $\tau_B$. 
Note that, if we make $\Delta t$ smaller and smaller (keeping other parameters fixed), the system dynamics according to the dynamical map $\Lambda_{\Omega,t}^S$ becomes practically a noiseless unitary evolution. Consequently, it will be futile to discuss the effects of noise correlations in the regime where there is almost no noise in the first place. To deal with this situation, we have to make sure that no matter how small $\Delta t$ we take, the amount of noise remains the same for a particular amount of sensing time (we assume that the control operations are instantaneous). 

To make that work, first we fix all the parameters and $\Delta t$ at certain initial values. With this, 
we have the initial amount of noise in the system determined by the decay in the population of system density matrix, which in turn is determined by $c_1(t)$ in Eq. (\ref{S-dynamical-map}). So initially, we have,
\begin{equation}
\label{rescaling}
    c_1(t_i)\equiv c_1(t_i, \lambda_i, \gamma_{0,i})
\end{equation}
where $t_i$ is the initial $\Delta t$ and similarly for $\lambda_i$ and $\gamma_{0,i}$. Now, let's say, we make $\Delta t$ smaller to a final value of $t_f$. This means the system will now evolve $n=t_f/t_i$ times for a single evolution corresponding to $\Delta t=t_i$. At each time evolution the system is coupled a fresh environment. In order to keep the `total noise fixed' we want to impose the condition that  ${(\Lambda_{\Omega,t_f}^S)}^n[\cdot]=\Lambda_{\Omega,t_i}^S[\cdot]$.
If we keep $\lambda$ fixed, the noise will remain same if we change the parameter $\gamma_{0,i}$ to a new $\gamma_{0,f}$, such that,
\begin{equation}
    c_1(t_i, \lambda_i, \gamma_{0,i})=c_1^n(t_f, \lambda_i, \gamma_{0,f})
\end{equation}
Note that, we could also keep $\gamma_0$ fixed and vary $\lambda$ to satisfy the equation, but as $\lambda$ is related to the bath correlation time, it makes sense to keep it unchanged to analyze the effects environmental-correlations as a function of  varying $\Delta t$. In Fig. \ref{fig:JC-model-plot-2}, we compared the evaluated QFI for the models considered above with a single qubit ancilla space. As mentioned before, we keep $\lambda$ fixed and vary $\Delta t$ and $\gamma_0$ accordingly, to keep the amount of noise same for a fixed sensing time irrespective of how many controls we apply in between. This construction of rescaling the parameter according to different $\Delta t$ implies that in the limit of $\Delta t\rightarrow 0$, the dynamics $\Lambda_{\Omega,\Delta t}^S$ is effectively Markovian and can be described by a master equation similar to Eq. (\ref{master}) but with a constant decay rate (say $\gamma_c$) instead of $\gamma(t)$, where $\gamma_c$ can be evaluated as the solution of the following equation,
\begin{equation}
\label{rescale-Markov}
    e^{-xt_i}=|c_1(t_i)|^2.
\end{equation}
In the above, $c_1(t_i)$ and $t_i$ are same as Eq. (\ref{rescaling}).
Consequently, the Markovian bound in Ref. \cite{Arpan2025} should be valid in this limit. To show this,
we add the Markovian bound curve that correspond to $\gamma_c = 1.14$, which is the solution of Eq. (\ref{rescale-Markov}) for 
$t_i =\Delta t =0.5$ and corresponding $\gamma_0=2.457$ (with fixed $\lambda\sim\tau_B^{-1}=2.5)$. Clearly, this provides a valid upper bound to the QFI computed for the \textit{fresh-environment} model but not for the correlated one. This shows the crucial role of the memory about the system, carried through the system-environment correlation in the estimation problem. One can also notice an interesting characteristic in the plots for case of fresh environments. As we keep decreasing $\Delta t$, the QFI  increase  
but then saturates very quickly. Understandably, in the limit of $\Delta t\rightarrow0$, the \textit{fresh-environment} model approaches to a Markovian amplitude damping dynamics with a constant decay rate $\gamma_c$, as explained before. On the other hand, in the model that includes environmental correlations, we see significant improvement, when we perform controls over time that are becoming shorter than the bath correlations times. 
This indicates, that by more frequent application of quantum controls we indeed `recover' more information that is being transmitted effectively through the state of the environment over the course of the protocol.  
The only case, where there is no obious discrpeancy between the two approaches is $\Delta t=2$, where the spearation between intergaion steps is long enough, so that both models are equivalent, as there is barely any correlation left in the environment on this time scale.
%This can be explained from the fact that as we keep on decreasing $\Delta t$, for a single run, the noise gets so small that it barely affects the initial state. As a result, the control operations in between can not do much to change the state in a way that increases the %QFI significantly. 
 
%Note that, if we keep $\lambda$ and $\gamma_0$ both constant, of course in the limit $\Delta t\rightarrow0$, the QFI will approach the unitary encoding limit (Heisenberg scaling) as there will be no noise. But in this analysis, we always keep the noise level constant (for a fixed sensing time) while comparing the results.
 
\section{Conclusion}
\label{section-5}
In this paper, we have provided a general framework of quantum adaptive metrological protocols for inherently quantum correlated noise coming from generic open quantum system, potentially non-Markovian and strongly coupled to the bath. Except for some special cases, for a generic open quantum system, exact system dynamics is basically non-existent unless one uses some approximations to cast it in the standard Markovian master equation form. This makes it incredibly hard to device a general setup for quantum metrology where quantum correlated noise is present. We circumvent this problem using a technique of Markovian embedding, where the non-Markovian system dynamics is mapped to an enlarged system involving finite and discrete degrees of freedom (called pseudomodes) obeying a Markovian master equation, such that the reduced system dynamics matches with the original one. In this paper, we have used this technique for a bosonic envuronment to which the system is coupled linearly. Now that we have master equation description of a dynamics with finite degrees of freedom, the pseudomodes can be treated as the inaccessible environment in an adaptive metrological protocol, where the parameter to be estimated is encoded in the system. The system-environment correlation is now present throughout the protocol serving as a memory, which, intuitively, should help to enhance the estimation precision. To exemplify this, we have provided an example of spontaneous decay model of two level system in a bosonic environment with zero temperature. This model has an exact solution for excitation preserving interaction Hamiltonian, which provides a way to compare the effects of correlated dynamics on quantum adpative metrological protocols with the uncorrelated one. We have seen a significant gain in the output QFI for the correlated environment model, evaluated numerically with ISS and tensor network techniques. For this model, we have also evaluated the recent correlated bounds, which holds universally irrespective of the type of noise and showed that it gives a legitimate upper bound for the QFI with correlated environment. Interestingly, our previous bound for Markovian quantum metrology only holds only when there is no correlation, which is expected.

We expect that the formalism outlined in this paper can be applied to more complicated quantum dynamics in a similar way. This can lead to a unified study of quantum metrology for inherently quantum correlated noise.

\begin{acknowledgments}
We thank Dariusz 
Chru{\'s}ci{\'n}ski for fruitful discussions
as well as Piotr Dulian and Staszek Kurdzia{\l}ek for the support in running the numerics with the help of the QMetro++ package. This work was supported by National Science
Center (Poland) grant No.2020/37/B/ST2/02134. AD acknowledges the support from the Grant RYC2022-
036958-I funded by MICIU/AEI/10.13039/501100011033
and by ESF+.
\end{acknowledgments}

\appendix
\section{Detail about Optimal control operations}
\label{appen-A}
Here, we describe the optimal control operations for the models described in Fig. \ref{fig:JC-model-plot}(a), for $t/\Delta t=N=2$ without ancilla. Protocols involving qubit ancilla can also be described in a similar way but for an intuitive understanding, the case with no ancilla is better suited. 

As mentioned earlier, the \textit{fresh-environment} model is a spontaneous decay model of a two-level system with time dependent decay rate as described in Eq. (\ref{master}), and so is the traced-out dynamics of the system for the correlated environment model as depicted in Eq. (\ref{S-dynamical-map}). As expected, for $N=1$, the optimal system input state is any state on the equator of the Bloch sphere: $\frac{1}{\sqrt{2}}(\ket{0}+e^{i\phi}\ket{1})$, like the spontaneous decay model with time-independent decay rate. Here $\ket{0}$ is the ground state of the Hamiltonian ($\frac{\Omega}{2}\sigma_z$) and denotes the south pole of the Bloch sphere.

For $N=2$, however, the optimal system input state of the protocol is no longer a state on the equator. For our example, with $\Delta t=0.5$, $\tau_B=0.4$ and $\gamma_0=2.457$, the optimal input system state for the fresh environment model, obtained from the numerics is: $\ket{\phi}^f_{\rm inp}=0.791853\ket{0}+(0.523167-0.315064i)\ket{1}$. Whereas, for the correlated environment model, the optimal input system state is $\ket{\phi}^c_{\rm inp}=0.758672\ket{0}+(0.36717-0.538147i)\ket{1}$. 
From the representation of density matrix as $\rho=\frac{1}{2}(\mathds{1}+\hat{n}.\vec{\sigma})$, with $\vec{\sigma}\equiv \{\sigma_x,\sigma_y,\sigma_z\}$, the unit vectors $\hat{n}\equiv\{n_x,n_y,n_z\}$ for these states are $\{0.828,0.499,-0.254\}$ and $\{0.557, 0.816, -0.151\}$ respectively. As we can see the Bloch vectors of these states are slightly below the equator.  
Now, if we rotate the states around $z$-axis it does not affect the optimality of the states but we can get rid of the azimuthal angle, such that the unit vectors lie in the $x-z$ plane. By doing that we get simpler representation of the states as,
\begin{align}
\label{inpf}
    &\ket{\phi}^f_{\rm inp}=0.791853\ket{0}+0.610712\ket{1}\\
    \label{inpc}
    & \ket{\phi}^c_{\rm inp}=0.758672\ket{0}+0.651473\ket{1}
\end{align}
The Bloch vectors of these states are $\{0.967, 0.0, -0.254\}$ and $\{0.988, 0.0, -0.151\}$ respectively. 
Recall that, the initial state of the environment $M$ is the ground state $\ket{0}_M$ of the Hamiltonian $H_M$. 
After the first channel acts on the fresh environment model according to $\Lambda^S_{\Omega,\Delta t}$, the noisy part of the dynamics makes the the $z$ component of the Bloch vector move towards the south pole and simultaneously shrinking the $x$ and $y$ components (this makes the state mixed). The Hamiltonian or the parameter encoding part rotates the Bloch vector around $z$ axis. For the correlated environment scenario, the channel $\Lambda^{SM}_{\Omega, \Delta t}$ acts both on the system and environment ($M$) space jointly. So, the effect on the system state is less obvious (as the tracing out of the environment is done at the end of the protocol before measurements) but one can observe the similar effects on the reduced system state.

Before discussing the control operations, we closely examine Eq. (\ref{S-dynamical-map}) to gain some intuition about what affects the QFI. For fixed $t$ (in our case $\Delta t=0.5$), the system state is of the following form (in Schr\"{o}dinger picture),
\begin{equation}
    \rho_S(t)= \begin{pmatrix}
        |c_1(t)|^2 & c_0^* c_1(t)e^{-i\Omega t}\\
        c_0 c_1^*(t)e^{i\Omega t} & 1-|c_1(t)|^2,
    \end{pmatrix}
\end{equation}
Note that, in the example we considered, we have $\Omega=\omega_0$, which implies that $c_1(t)$ and also its derivative are independent of $\Omega$ (this is also true even if they are not zero). So, $\Omega$ dependence only comes from the off-diagonal elements. One can directly calculate the QFI for this state:
\begin{equation}
    F_Q(t)=4|c_0^* c_1(t)|^2t^2
\end{equation}
This shows that QFI is directly proportional to the coherence of the state, which is given as ($l_1$ norm of coherence) $\mathcal{C}=2|c_0^* c_1(t)|$. From the Bloch vector representation of density matrix, coherence is nothing but the transverse radius $\sqrt{n_x^2+n_y^2}$ of the unit vector $\hat{n}$. As discussed before, the effect of the dynamics on the state is reducing the transverse radius and thus moving the $z$ component more towards the south pole. Naturally, the control operation should be such that it increases the transverse radius and pushes the $z$ component of the Bloch vector towards the equator. Indeed, from the numerics (with initial states as Eq. (\ref{inpf} and Eq. (\ref{inpc})) we get two unitaries: $U_f$ for the fresh environment model and $U_c$ for the correlated environment model,
% \begin{align}
%     \label{unitraies}
%     &U_f= \begin{pmatrix}
%         0.908+0.257i & 0.318-0.087i\\
%         -0.283-0.170i & 0.944
%     \end{pmatrix},\\
%     &U_c=\begin{pmatrix}
%         -0.974+0.023i & -0.124+0.188i\\
%         -0.129-0.185i & 0.974
%     \end{pmatrix}.   
% \end{align}
which exactly do this job. To describe these unitaries, note that, any $(2\times 2)$ unitary (which has $4$ real independent parameters) can be expressed as $e^{i\alpha} e^{-i\theta\hat{n}.\vec{\sigma}/2 }$. But in our case (rotation of Bloch vectors), the phase factor $e^{i\alpha}$ does not matter. With that, $U_f$ rotates the Bloch vector of the state by an angle $0.22\pi$ about an axis: $\{-0.072, -0.976, -0.206\}$ and $U_c$ rotates it by an angle $0.99\pi$ about an axis: $\{-0.225, -0.0007, -0.974\}$. Note that the axis of the rotation for $U_f$ almost lies in the $y-z$ plane, whereas the axis of rotation for $U_c$ almost lies in the $x-z$ plane.
Recall that the parameters of the problem: $\Delta t=0.5$, $\tau_B=0.4$, $\gamma_0=2.457$, and $N=2$.
The unitaries are optimized in a way that after the application of second and last channel it ensures that the output states (traced out state for correlated environment model) have the maximum coherence possible. Thanks to the environment correlation for the second model, which always produces a state with larger coherence and hence larger QFI than the fresh environment model. 
Note that, the choice of control unitary is necessary for this conclusion. For example, instead of the optimal unitary $U_c$ as given by the numerics, had we used the same unitary $U_f$ as the fresh environment model, we would end up with less coherent state and hence less QFI than in the fresh environment model. 

Now, if we decrease $\gamma_0$ (or decrease $\tau_B$), keeping everything else fixed, it effectively reduces the amount of noise (decay in the density matrix) in the dynamics. Consequently, the system state gets less mixed after the application of the channel and control unitaries become more effective to produce final states with better coherence and hence increasing the QFI. For example, we have checked that when $\gamma_0$ is reduced from $2.457$ to $1.5$, keeping $\tau_B=0.4$ and $\Delta t=0.5$ fixed, for $N=2$, output QFI is increased by about $23\%$ for the correlated case and by about $32\%$ for the fresh environment case. The optimal initial states for both scenarios are more close to the equator compared to the previous parameter settings.
Even if we apply the same control unitaries for the previous parameter settings (which are not optimal for the present parameter values), we achieve almost optimal QFI in both scenarios. On the other hand if we decrease $\Delta t$, keeping everything else fixed, the duration of the dynamics for each run is decreased resulting less accumulation of noise. Consequently, for same sensing time, more control unitaries are employed to rectify less affected system states, which in turn increase the QFI. For example, in the previous scenario for $\Delta t=0.5$ and $N=2$, total sensing time is $1$. If we decrease $\Delta t$ to $0.25$, for the same sensing time we now have $N=4$. As a result, QFI increases by about $42\%$ for the correlated and by $33\%$ for the fresh environment scenario.  

\bibliography{timescales}

\end{document}